\documentclass[amssymb,aps,prl,twocolumn]{revtex4-1}

\usepackage{amsmath,color}
\usepackage{amssymb}
\usepackage{amsfonts}

\usepackage{graphicx}
\usepackage{framed}

\begin{document}

\title{Heterodyne-Detected Ultrafast X-Ray Diffraction and Scattering from Nonstationary States}
\author{Kochise Bennett$^{a,b}$}
\email{kcbennet@uci.edu}
\author{Markus Kowalewski$^{a}$}
\author{Shaul Mukamel$^{a,b}$}
\email{smukamel@uci.edu}
\affiliation{$^a$Chemistry Department, University of California, Irvine, California 92697-2025, USA}
\affiliation{$^b$Department of Physics and Astronomy, University of California, Irvine, California 92697-2025, USA}

\date{\today}
\begin{abstract}
Free-electron laser hard X-ray light sources can provide high fluence, femtosecond pulses, enabling the time-resolved probing of structural dynamics and elementary relaxation processes in molecules. Traditional X-ray elastic scattering from crystals in the ground state consists of sharp Bragg diffraction peaks that arise from pairs of molecules and reveal the ground state charge density. Scattering of ultrashort X-ray pulses from gases, liquids, and even single molecules is more complex and involves both single- and two- molecule contributions, diffuse (non-Bragg) features, elastic and inelastic components, contributions of electronic coherences in nonstationary states, and interferences between scattering off different states (heterodyne detection). We present a unified description that covers all these processes and discuss their relative magnitudes for gas-phase NaI.  Conditions for the observation of holographic (heterodyne) interference, which has been recently discussed  \cite{glownia2016self}, are clarified.
% X-ray scattering in crystals is dominated by elastic, 2-molecule contributions. This in not the case for translationally disordered systems in solution. Diffraction signals from electronically excited states interfere with the stronger scattering from unexcited molecules in the ground state to create a holographic heterodyne detection scheme. We derive a unified expression for X-ray scattering signals that includes single- and two-molecule contributions, elastic and inelastic terms, and nuclear dynamics for a molecule prepared in an arbitrary superposition state including electronic coherence.
\end{abstract}
\maketitle
%OLD ABSTRACT: X-ray diffraction has long been employed to discover the real-space structure of crystals on atomic length scales.  More recently, extension has been made to apply free-electron lasers to single-molecules with the hopes of avoiding the crystalization process as well as to time-resolved studies with the goal of observing structural dynamics in real time.  These extensions raise some basic questions about how the time-resolved diffraction signal should be defined and calculated.  In particular, the difference between two-molecule and single-molecule diffraction is seldom addressed.  In this manuscript we clarify these issues, deriving a unified expression that includes single- and two-molecule contributions, elastic and inelastic terms, and nuclear dynamics for a molecule prepared in an arbitrary state including electronic coherence.  
%\section{introduction}
The term diffraction refers to the off-resonant elastic scattering of light \cite{guiner, modxrayphys,velser}.  From a simple, classical picture, the amplitude of the light scattered from each location in the material acquires a spatial phase-factor and repeated spatial patterns lead to Bragg peaks in the scattered signal where the light scattered from different points in the sample adds coherently.  This technique has been used for over a century to probe the structure of crystals and has been extended to diffuse scattering from liquids, probing nearest-neighbor distances and served as inspiration for similar the conceptually similar electron diffraction technique \cite{ben1996direct,siwick2003atomic}.  More recently, effort has been made to push diffraction to the single-molecule limit, eliminating the need for time-consuming crystal preparation \cite{stevens, mcpherson, hajdu, chapmanbeyond, starodub}.  Time-resolved X-ray diffraction is a natural way to track the structural changes that characterize phase transitions and chemical reactions and has been actively pursued to create molecular movies \cite{bratos2002time,siwick2003atomic,coppens2005structure,ihee2005ultrafast, wulff2006recombination, cammarata2008tracking,siders,woerner2010concerted, coppens2011molecular, neutze2012time, falcone}. These efforts have been made possible by the development of free-electron hard X-ray sources capable of producing bright, femtosecond-duration pulses \cite{altarelli, feldhaus, mcneil, naturefsnano,bostedt2013ultra,barty2013molecular}.
\par
Heterodyne detection involves the interference of a weak optical signal field with a strong reference (local oscillator).  The resulting signal is linear (rather than quadratic) in the weaker signal field and thus scales favorably in addition to revealing phase information.  Such holographic detection is well established in the visible regime and has been extended to transient X-ray diffraction in crystals and powders \cite{vrakking2012x, elsaesser} and was recently discussed in the gas phase \cite{glownia2016self}.  For weak excitations, where only a small fraction of the molecules is excited, the signal from the ground state molecules serves as a local oscillator for the weaker excited state signal.  An external local oscillator is not needed since it is generated \textit{in situ}.  This amounts to self-heterodyne detection. 
\par
X-ray diffraction from crystals in the ground state is purely elastic, contains no electronic coherence, and is given by a product of scattering amplitudes of two molecules.  Time-resolved scattering from photoexcited molecules in the gas phase, in contrast, is a sum of single-molecule contributions, contains elastic and inelastic contributions, and can depend on electronic coherence.  We calculate the X-ray scattering by an ensemble of molecules prepared in a superposition of valence electronic states and identify the various contributions and show that, in the absence of valence electronic coherence and inelastic X-ray scattering, the gas-phase diffraction signal is simply given by the sum of ground- and excited-state contributions and contains no cross (heterodyne) terms.
\par
%Heterodyne detection was reported in the recent gas-phase measurements of molecular iodine \cite{glownia2016self}.  analysis of Ref.\ \cite{glownia2016self}.
The total charge-density operator for a system composed of molecules can be written as a sum of the charge densities from each molecule
\begin{align}
\hat{\sigma}_\text{T}(\mathbf{r})=\sum_\alpha\hat{\sigma}_\alpha(\mathbf{r}-\mathbf{r}_\alpha)=\sum_\alpha\int d\mathbf{q} e^{i\mathbf{q}\cdot(\mathbf{r}-\mathbf{r}_\alpha)}\hat{\sigma}_\alpha(\mathbf{q})
\end{align}
where $\mathbf{r}_\alpha$ is the center of molecule $\alpha$.  For a sufficiently dilute system such that the molecules have non-overlapping charge distributions, this separation is exact since each electron (the fundamental X-ray scatterer) can be rigourously assigned to a specific molecule.  More generally, there will ordinarily be very little intermolecular electron density and this separation is justified.  For identical molecules, the charge density operators of each molecule differ only by the spatial phase-factor associated with the location of the molecule and we may drop the $\alpha$ subscript.
Elastic light scattering comes proportional to the Thomson scattering cross section which gives the intensity distribution of free-electron scattering \cite{guiner, modxrayphys}.  Neglecting this and other pre-factors, the diffraction signal from a system initially in the ground state $\vert g\rangle$ is
\begin{equation}\label{eq:SkClassic}
S(\mathbf{q})=\vert\sigma_{gg}(\mathbf{q})\vert^2,
\end{equation}
where $\sigma_{gg}(\mathbf{q})=\langle g\vert \hat{\sigma}(\mathbf{q})\vert g\rangle$ is the ground state charge
density in $\mathbf{q}$-space where $\mathbf{q}$ is the momentum transfer. Equation (\ref{eq:SkClassic}) assumes that the scattering is elastic.  In a previous work \cite{bennett2014time}, we derived expressions for 1-molecule and 2-molecule contributions to frequency-resolved diffraction which we denote as $S_1$ and $S_2$ respectively \footnote{Since the 2-molecule contributions carry spatial phase-factors that require long-range coherent order to observe, they had been termed coherent in Ref.\ \cite{bennett2014time}.  The single-molecule contributions on the other hand were termed incoherent since they add incoherently for the total signal.  Nonetheless, the 1-molecule contribution still has intramolecular spatial coherences (between atoms) so for clarity we use in this work the terms 1- and 2-molecule rather than incoherent and coherent}. In the supplementary information, we integrate out this frequency-resolution and take a quasi-elastic limit to arrive at the following simpler formulas
\begin{align}\label{eq:Scohtr2}
S_{2}(\mathbf{q},t)&=F(\mathbf{q})\int dt \vert E_p(t)\vert^2 \vert\langle\hat{\sigma}(\mathbf{q},t)\rangle\vert^2 \\ 
S_{1}(\mathbf{q},t)&= N\int dt  \vert E_p(t)\vert^2\langle\vert\hat{\sigma}(\mathbf{q},t)\vert^2\rangle \label{eq:Sinctr2}
\end{align}
where $E_p(t)$ is the temporal envelope of the X-ray pulse, $\langle\dots\rangle$ stands for expectation value over the nuclear and electronic states, and the function
\begin{align}
F(\mathbf{q})&=\sum_\alpha\sum_{\beta\ne\alpha}e^{-i\mathbf{q}\cdot(\mathbf{r}_\alpha-\mathbf{r}_\beta)}
\end{align}
is known as the structure factor and encodes the long-range (intermolecular) order of the sample, the intramolecular structure being encoded in the $\hat{\sigma}$. Note the subtle distinction between these two expressions: Eq.\ (\ref{eq:Scohtr2}) comes with $\vert \langle\hat{\sigma}(\mathbf{q}\rangle\vert^2$ while Eq.\ (\ref{eq:Sinctr2}) comes with $\langle\vert\hat{\sigma}(\mathbf{q})\vert^2\rangle$.  The former coincides with the classical definition of diffraction (Eq.\ (\ref{eq:SkClassic})) but is actually due to the coherent addition of the scattering amplitude from pairs of molecules while the latter accounts for single-molecule diffraction.  In a crystal, the long-range order gives rise to sharp Bragg peaks in $F(\mathbf{q})$ while the 1-molecule terms form a diffuse background that can be neglected.  The enhancement at these Bragg peaks scales quadratically in the molecule number (i.e., $\propto N^2$) and the intensity at each peak comes proportional to the $\mathbf{q}$-space charge density at that point.  This is the traditional picture of diffraction. Sampling the molecular $\mathbf{q}$-space charge density at the Bragg peaks is sufficient to reconstruct the magnitude of the $\mathbf{q}$-space charge density.  By considering the diffuse scattering between Bragg peaks, enough information can be obtained to reconstruct the phase as well, providing one solution to the phase problem in X-ray diffraction \textit{via} oversampling \cite{miao1999extending,robinson2001reconstruction,miao2003phase}. 
\par
In the continuum limit, we can replace the summations over molecules with spatial integrations.  Assuming spatial homogeneity (no long-range order), we obtain $S_2\propto\delta(\mathbf{q})$ and the 2-molecule terms contribute only at zero momentum transfer (the central maximum of the diffraction pattern) and the 1-molecule term becomes dominant. Thus, in the absence of long-range order, such as in a gas, the diffraction signal should be simulated with Eq.\ (\ref{eq:Sinctr2}) rather than (\ref{eq:Scohtr2}).  It is a common error to write the single-molecule diffraction as Eq.\ (\ref{eq:Sinctr2}) but with $\vert\langle\hat{\sigma}\rangle\vert^2$ rather than $\langle\vert\hat{\sigma}\vert^2\rangle$. When all molecules in the sample are in the ground state and inelasticities are ignored, these two are the same, both being $\vert\sigma_{gg}(\mathbf{q})\vert^2$.  More generally however, they result in different types of terms that, as will be shown next, affect the interpretation of the diffraction signal from nonstationary states.

\begin{figure}
\includegraphics[width =.95\linewidth]{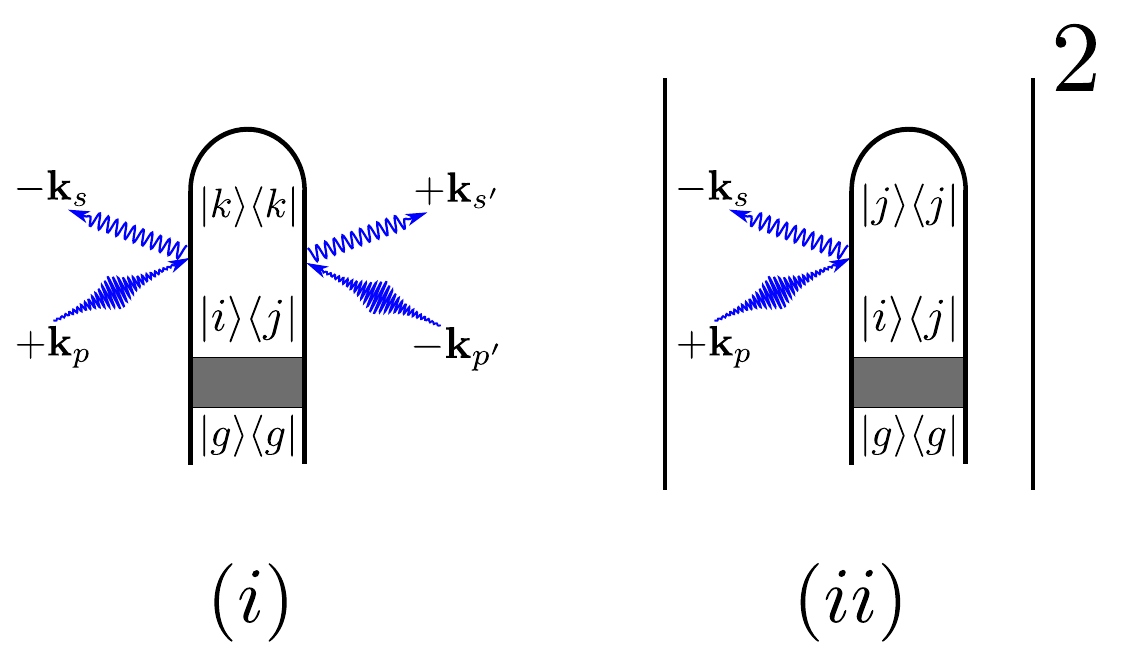}
\caption{Loop diagrams for single-molecule (i) and two-molecule (ii) X-ray scattering processes. The amplitude-squared form of the two-molecule contribution has been explicitly indicated. The shaded area represents an excitation process that prepares the system in an arbitrary superposition state ($\vert g\rangle$ is the electronic ground state).  We denote modes of the X-ray pulse with $p$ and $p'$ whereas $s$, $s'$ represent relevant scattering modes ($\mathbf{k}_{p^{(\prime)}}$ has frequency $\omega_{p^{(\prime)}}$ and $\mathbf{k}_{s^{(\prime)}}$ has frequency $\omega_{s^{(\prime)}}$). Note that we use $\vert\phi_i\rangle\to\vert i\rangle$ for the electronic states in this figure to aid readability. A complete list of diagrams is given Fig.\ (S1).}
\label{fig:diagexp}
\end{figure}

%\subsection{X-Ray Scattering with Electronic Coherence and Nuclear Dynamics}%(Version A)
We consider a molecular model consisting of two electronic states $e,g$ and a single active nuclear coordinate $\mathbf{R}$.  The time-dependent wavefunction of each molecule in the ensemble will be expanded in the product space
%\begin{align}
%\vert \Psi\rangle=\sum_{i=e,g}c_i\vert\chi_i\rangle\otimes\vert \phi_i\rangle
%\end{align}&=\sum_{i=e,g}c_i(t)\vert\chi_i(t)\rangle\otimes\vert \phi_i\rangle\\ \notag &
\begin{align}
\vert \Psi(t)\rangle=e^{-i\hat{H}_0t}\sum_{i=e,g}c_i(0)\vert\chi_i(0)\rangle\otimes\vert \phi_i\rangle
\end{align}
where $\vert\chi_i\rangle$ is the (normalized) nuclear wavepacket on electronic state $\vert\phi_i\rangle$, $c_i$ is the electronic state amplitude, and $\hat{H}_0$ is the field-free nuclear Hamiltonian.  For a weak excitation, we will have $\vert c_e\vert^2=\epsilon\ll1$ and $\sum_i\vert c_i\vert^2=1$.  We treat the system in the Born-Oppenheimer approximation (BOA) wherein the nuclear wavepackets on each electronic eigenstate evolve independently and the Hamiltonian seperates into the sum of kinetic and potential energies on each $\vert\phi_i\rangle$.
Explicitly expanding the results given in the supplement in the electronic states gives
\begin{widetext}
\begin{align}\label{eq:S1molExp}
S_1(\mathbf{q},t)=N\bigg\lbrace\rho_{gg}(t)\langle\chi_g(t)\vert \hat{\sigma}^\dagger_{gg}(\mathbf{q})\hat{\sigma}_{gg}(\mathbf{q})&+\hat{\sigma}^\dagger_{ge}(\mathbf{q})\hat{\sigma}_{eg}(\mathbf{q})\vert\chi_g(t)\rangle + \rho_{ee}(t)\langle\chi_e(t)\vert \hat{\sigma}^\dagger_{ee}(\mathbf{q})\hat{\sigma}_{ee}(\mathbf{q})+\hat{\sigma}^\dagger_{eg}(\mathbf{q})\hat{\sigma}_{ge}(\mathbf{q})\vert\chi_e(t)\rangle \\ \nonumber
+&2\Re\big[\rho_{eg}(t)\langle\chi_e(t)\vert\hat{\sigma}^\dagger_{ee}(\mathbf{q})\hat{\sigma}_{eg}(\mathbf{q})+\hat{\sigma}^\dagger_{eg}(\mathbf{q})\hat{\sigma}_{gg}(\mathbf{q})\vert\chi_g(t)\rangle\big]\bigg\rbrace
\end{align}
\begin{align}\label{eq:S2molAmp}
S_2(\mathbf{q},t)=&F(\mathbf{q})\bigg\vert \rho_{gg}(t)\langle\chi_g(t)\vert\hat{\sigma}_{gg}(\mathbf{q})\vert\chi_g(t)\rangle+\rho_{ee}(t)\langle\chi_e(t)\vert\hat{\sigma}_{ee}(\mathbf{q})\vert\chi_e(t)\rangle+2\Re\big[\rho_{eg}(t)\langle\chi_e(t)\vert\hat{\sigma}_{eg}(\mathbf{q})\vert\chi_g(t)\rangle\big]\bigg\vert^2.
\end{align}
\end{widetext}
where the electronic populations and coherences are given by the diagonal and off-diagonal elements of the density matrix $\rho_{ij}(t)=c^*_i(t)c_j(t)$ and we have defined the electronic-state matrix elements of the charge-density operator $\hat{\sigma}_{ij}(\mathbf{q})=\langle\phi_i\vert\hat{\sigma}(\mathbf{q})\vert\phi_j\rangle$ (which remains an operator in the nuclear space) and, for brevity, omitted the time integration over the X-ray pulse envelope.
% In a quantum-electrodynamic (QED) description, the density matrix of the scattered field mode goes from $\vert 0\rangle\langle 0\vert$ to $\vert 1\rangle\langle 1\vert$ in photon number states.
%but the latter two such terms differ from the first two by $e\leftrightarrow g$ and the two sets are complex conjugates that we re-write using the real part for brevity.Note that these scatterings off electronic coherences contain heterodyne beatings between the ground (excited) state charge density $\sigma_{gg}$ ($\sigma_{ee}$) and the transition state charge density and may thus complicate efforts to detect the heterodyne interference between $\sigma_{ee}$ and $\sigma_{gg}$, at least prior to decoherence. 
\par
The first four terms on the right-hand side of Eq.\ (\ref{eq:S1molExp}) represent the elastic ($\sigma_{gg}$ and $\sigma_{ee}$) and inelastic ($\sigma_{eg}^{(\dagger)}$) scattering from the electronic ground and excited state populations. The final two terms of Eq.\ (\ref{eq:S1molExp}) are due to scattering off coherences (we have used the fact that terms related by $e\leftrightarrow g$ are complex conjugates to simplify).  In contrast to the 1-molecule signal, the 2-molecule signal is given by the modulus square of an amplitude. The first two terms in this amplitude (Eq.\ (\ref{eq:S2molAmp})) correspond to the ground- and excited-state scattering amplitudes respectively while the final amplitude term is the scattering from coherences.  We note that 2-molecule scattering from populations is purely elastic while 1-molecule scattering from populations contains both elastic and inelastic terms.  In both cases, the presence of electronic coherences introduces new terms that result in heterodyne interference between transition charge densities and population charge densities, though the precise form of this contribution varies in the two cases.
\par%For this reason, we propose to reserve the term ``diffraction'' for 2-molecule elastic X-ray scattering, and refer to more general sum of 1- and 2-molecule contributions by the generic term ``X-ray scattering''. 
The diffraction signal is commonly taken to be $S\propto\vert\langle\hat{\sigma}\rangle \vert^2$.  This is correct for the 2-molecule contribution but does not generally hold for the 1-molecule contribution which is given by $S_1\propto\langle\hat{\sigma}^\dagger\hat{\sigma}\rangle$.  There are two important but subtle differences between these expressions: (1) heterodyne interference between ground- and excited-state scattering (i.e., terms of the form $\sigma_{ee}\sigma_{gg}$) appear only in $S_2$ and (2) $S_2$ is proportional to $\rho^2$ while $S_1$ is linear in $\rho$ so that the excited state diffraction $\sigma_{ee}$ comes proportional to $\rho_{ee}$ for $S_1$ rather than $\rho_{ee}^2$ for $S_2$.   Since the 2-molecule contribution scales quadratically with the molecule number $N$ at the Bragg peaks, this signal overwhelms the 1-molecule scattering there.  In between the Bragg peaks, or in the absence of long-range order, the signal is governed by the 1-molecule scattering and should be calculated using Eq.\ (\ref{eq:S1molExp}) (note the similarity to the discussion in Ref.\ \cite{guiner}, \S 3.1.3). The 1-molecule contribution is then linear in the molecular density matrix $\rho$ and identified with Raman scattering while the 2-molecule is quadratic in $\rho$ and corresponds to Rayleigh scattering \cite{dorfman}. While we are not the first to point out that diffraction from nonstationary states differs from a simple $\vert\sigma(\mathbf{q})\vert^2$ form \cite{bratos2002time,schulke,modxrayphys,dixit}, it is still widely employed.  We suspect that the confusion on this point is exacerbated by a lack of clarity on the separation of terms into single- \textit{vs.\ }two-molecule.  While diffraction is often understood from an independent atom perspective and separated into single- and two-\emph{atom} terms, this approach treats intermolecular and intramolecular structure on the same footing, as distances between atoms in the sample. Since the molecular structure is the quantity of interest and molecular bonding electrons are not independent, it is more appropriate and accurate to treat the inter- and intra-molecular structure seperately as above.  Moreover, our approach includes inelasticities, leading to transition charge densities $\hat{\sigma}_{ij}(\mathbf{q})$ that are usually neglected but that interfere directly with ground and excited state terms $\hat{\sigma}_{ii}(\mathbf{q})$.
%should be carried out via $\langle\vert\hat{\sigma}(\mathbf{q})^2\vert\rangle$ rather than We note that heterodyne detection, which is intended to probe the interference between $\sigma_{gg}(\mathbf{q})$ and $\sigma_{ee}(\mathbf{q})$, is only applicable to the 2-molecule contributions as no such terms appear in the single-molecule scattering. In Ref.\ \cite{glownia2016self}, they neglected electronic coherences and inelastic scattering and assumed that $S_2$ averages out due to the lack of long-range order. According to our results, the signal is then given by the sum of $\rho_{gg}$ and $\rho_{ee}$ single molecule terms and there is no interference (heterodyne detection). The heterodyne detection is a macroscopic interference between the signals created by the two groups of molecules, excited and un-excited.  Interference within a molecule can only exist if we keep electronic coherences.
%\section*{Discussion}

\par
We demonstrate the relative magnitude of the different contributions to the single-molecule signal
in Eq. \ref{eq:S1molExp}, for sodium iodide. The two relevant valence states
are the X$^1\Sigma^+$ ground state and the A$^1\Sigma^+$ state (referred to as
$g$ and $e$ in the following).
\begin{figure}
\centering
\includegraphics[width=0.5\textwidth]{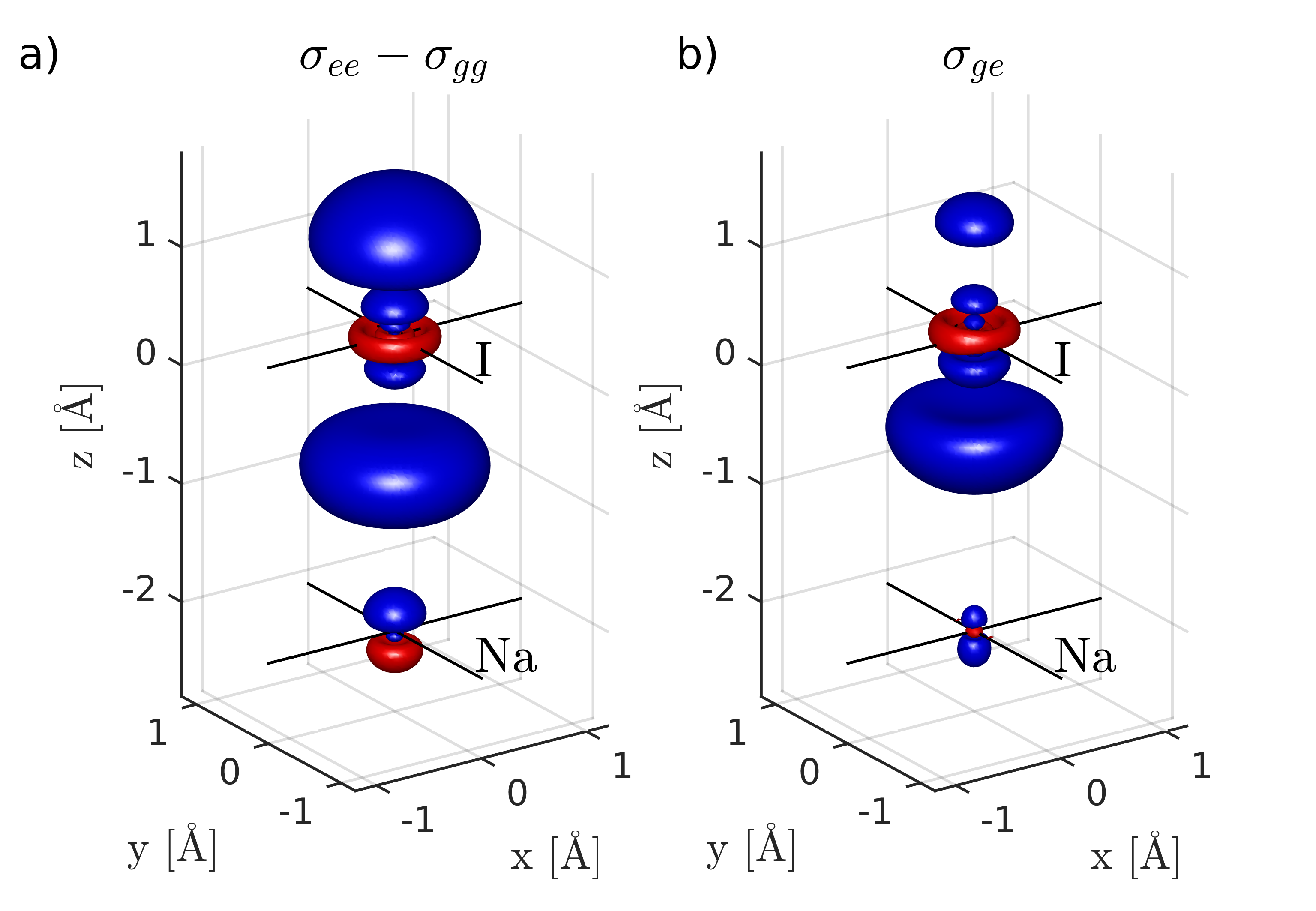}
\caption{Isosurfaces of the charge densities of NaI.  The difference density (a) $\sigma_{ee}-\sigma_{gg}$ and the transition density (b) $\sigma_{ge}$ are evaluated for isovalues of 0.01 and 0.005 respectively (red: positive sign, blue: negative sign).
The crosshairs mark the atomic centers.} \label{fig:sigma3d}
\end{figure}
Figure \ref{fig:sigma3d} shows the difference density between $\sigma_{gg}$ and $\sigma_{ee}$ as well as the transition density $\sigma_{ge}$. The $g\rightarrow e$ excitation is characterized by promoting an electron from $n_{p_z}$ of the iodine into a $\sigma^*$ bond, thus weakening the bond. This can be clearly seen in Fig. \ref{fig:sigma3d}(a):
electron density from in between the atoms and the lone pair at the iodine is removed. This
feature is represented in similar way in the transition density (Fig. \ref{fig:sigma3d}(b)).

\begin{figure}
\centering
\includegraphics[width=0.45\textwidth]{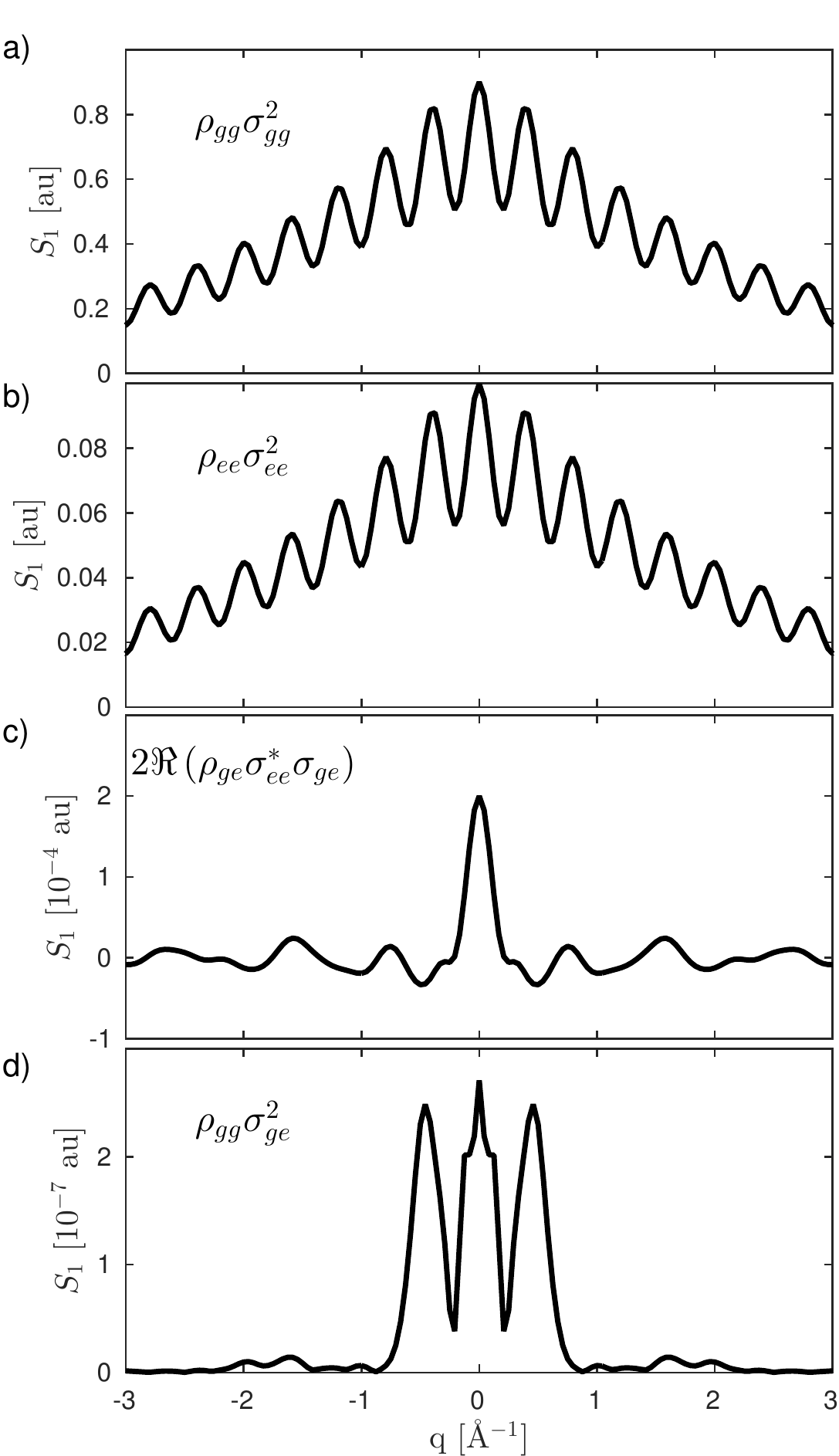}
\caption{Contributions to the single particle signal (see Eq. \ref{eq:S1molExp}). The components of the  signal show the projection onto the molecular axis and are normalized to $\sigma_{gg}^2(0)=1$.
The density matrix of the electronic states is given by $\rho_{gg}=0.9$, $\rho_{ee}=0.1$,
$\rho_{ge}=0.3$.}
\label{fig:S1}
\end{figure}
In Fig. \ref{fig:S1} the different contributions from the sum terms in Eq. \ref{eq:S1molExp} are shown for a fixed nuclear configuration ($R=2.5$\,\AA). Assuming an excitation fraction of 10\% ($\rho_{ee}=0.1$),
the strongest contribution to the $S_1$ signal stems from the ground state density ($\sigma_{gg}^2$) followed by the excited state density ($\sigma_{ee}^2$), which
is by one order of magnitude weaker (proportional to $\rho_{ee}$). The diffraction pattern allows to directly determine bond length.
In Fig. \ref{fig:S1}(c) the contribution which stems from valence electronic coherences
($\rho_{ge} \sigma_{ee}^*\sigma_{ge}$) combined with inelastic X-ray scattering
is shown for when the electronic coherence is maximal (e.g. directly after excitation with a pump). This contribution is $\approx$ 3-4 orders of magnitudes weaker than the portion of the signal, which stems from $\sigma_{ee}^2$ and is expected to rapidly decay as
the nuclear wave packet leaves the Franck-Condon region.
Contributions caused solely by the transition
densities ($\rho_{ge}\sigma_{ge}^2$) shown 
in Fig. \ref{fig:S1}(d), are $\approx$ 6 orders of magnitudes weaker
than the excited state density contribution.
This scaling behavior can be explained by the fact that diagonal densities ($\rho_{ee}$ and $\rho_{ee}$) are dominated by densely packed core electrons, while the transition densities are determined
by $\approx 1$ electron which is distributed over a valence orbital.

It becomes clear that $\sigma_{ee}^2$ -- assuming that the phase problem can be solved --  is sufficient to qualitatively recover the nuclear wave packet motion in the excited state. This part contains solely the phase of the excited state nuclear wave packet.
Given sufficiently short probe pulses (the pulse bandwidth must cover the energy gap between $e$ and $g$), the contributions from $\rho_{ge}\sigma_{ee}^*\sigma_{ge}$
can potentially be retrieved. This, temporally fast oscillating, part of the signal contains
the electronic phase information.

\par
Ultrafast diffraction from photoexcited iodine in the gas phase was recently reported \cite{glownia2016self}. By taking the 2-molecule contribution to vanish (due to the lack of long-range order in the gas sample) while neglecting electronic coherences and inelastic scattering, Eq.\ (\ref{eq:S1molExp}) finally gives just an incoherent sum of scattering from each molecule  $S_1\propto\rho_{ee}\vert\sigma_{ee}\vert^2+\rho_{gg}\vert\sigma_{gg}\vert^2$, i.e., the signal does look like that of a homogeneous mixture of excited and ground-state molecules. 
%\textcolor{red}{All of this clearly contradicts Eq.\ (4) of Ref.\ \cite{glownia2016self} and the discussion that follows.  The sentence ``The key insight in Eq.\ (4) is that scattering from the excited fraction in each molecule interferes with scattering from its initial state fraction, producing  holographic fringes.\'' is incorrect.  The molecule does not interfere with itself in the absence of electronic coherence.}
We find no heterodyne $\sim\sigma_{gg}\sigma_{ee}$ terms as claimed in Ref.\ \cite{glownia2016self}.  Such terms do exist in time-resolved Bragg peaks in crystals, which is a 2-molecule signal, but are absent from the incoherent sum of single-molecule terms that characterizes gas-phase signals. We remark that the difference in the $\rho$-scaling between the single- and two-molecule terms is crucial: if the sample is perturbatively pumped so that some small percentage $\rho_{ee}$ of molecules are in the excited state, 1-molecule scattering from the excited state is significantly stronger, compared to ground-state scattering, than it is in 2-molecule scattering ($\rho_{ee}$ \textit{vs} $\rho_{ee}^2$ respectively).  In fact, the 1-molecule excited-state scattering scales the same in $\rho$ as the 2-molecule holographic interference that is the object of heterodyne detection, opening the door to confusion.  Experimentally sorting out the various terms in the diagrams in Fig.\ \ref{fig:diagexp} will be an interesting future challenge. Finally, we note that homodyne versus heterodyne detection is a purely classical issue related to the macroscopic interference of light and has nothing to do with entanglement or Schroedinger cat states, as was incorrectly argued in Ref.\ \cite{buckpr}.  Quantum features can only be created by electronic coherences which were neglected in Ref.\ \cite{glownia2016self}.
%That is, terms of the form $\sigma_{gg}\sigma_{ee}$ only come from two-molecule contributions and would average to zero in the absence of long-range order as in a gas.
%Moreover, for a diagonal density matrix ($\rho_{eg}=\rho_{ge}=0$), the purely elastic single-molecule scattering contributions of the ground and excited state add incoherently. In contrast to the statements following Eq.\ (4) in Ref.\ \cite{glownia2016self}, the sample can then be viewed as an inhomogeneous distribution of two species in states $g$ and $e$, the fundamental meaning of diagonal density matrices.  This complicates the interpretation of the heterodyne detected X-ray diffraction technique to samples lacking long-range order.  

\acknowledgements 
The support of the Chemical Sciences, Geosciences, and Biosciences division, Office of Basic
Energy Sciences, Office of Science, U.S. Department of Energy through award No. DE-
FG02-04ER15571 as well as from the National Science Foundation (grant CHE-1361516) is gratefully acknowledged. Support for K.B. was provided by DOE. M.K. gratefully acknowledges support from the Alexander von Humboldt foundation through the Feodor Lynen program.

\appendix

%\clearpage

%\nocite{*}
%\bibliographystyle{plain}
\bibliography{TRXS,XRD}
\end{document}